\documentclass[twocolumn,prb,superscriptaddress,showpacs,preprintnumbers]{revtex4}

\usepackage{amsmath,graphicx,dcolumn}

\begin{document}

\preprint{LBNL-54103}

\title{Structural properties of the geometrically frustrated
pyrochlore Tb$_2$Ti$_2$O$_7$}

\author{S.-W. Han}
\altaffiliation{Present address: Chonbuk National University, Jeonju, 561-756, Korea}
\email{swhan@mail.chonbuk.ac.kr}
\affiliation{Chemical Sciences Division, Lawrence Berkeley National Laboratory,
Berkeley, California 94720}

\author{J. S. Gardner}
\email{jason.gardner@nist.gov}
\affiliation{
NIST Center for Neutron Research, National Institute of
Standards and Technology, Gaithersburg, Maryland 20899-8562}
\affiliation{Physics Department, Brookhaven National Laboratory, 
Upton, New York 11973}

\author{C. H. Booth}
\email{chbooth@lbl.gov}
\affiliation{Chemical Sciences Division, Lawrence Berkeley National Laboratory,
Berkeley, California 94720}

\begin{abstract}
Although materials that exhibit nearest-neighbor-only antiferromagnetic 
interactions and geometrical
frustration theoretically should not magnetically order in the absence of
disorder, few such systems have been observed experimentally.  One such
system appears to be the pyrochlore Tb$_2$Ti$_2$O$_7$.  However, previous
structural studies indicated that Tb$_2$Ti$_2$O$_7$ is an imperfect pyrochlore.
To clarify the situation, we performed neutron powder
diffraction (NPD) and x-ray absorption fine structure (XAFS) measurements 
on samples that were prepared identically to those that show no magnetic order.
The NPD measurements show that the long-range
structure of Tb$_2$Ti$_2$O$_7$ is well ordered with no structural 
transitions between 4.5 and 600~K.  In particular, mean-squared
displacements ($u^2$'s) for each site follow a Debye model with no offsets.  No
evidence for Tb/Ti site interchange was observed within an upper limit of 2\%.
Likewise, no excess or deficiency in the oxygen stoichiometry was observed, 
within an upper limit of 2\% of the nominal pyrochlore value.
Tb $L_\textrm{III}$- and Ti $K$-edge XAFS measurements from 20-300 K similarly 
indicate a well-ordered local structure.  Other aspects of the structure are
considered.  We conclude that Tb$_2$Ti$_2$O$_7$ has, within experimental error,
an ideal, disorder-free pyrochlore lattice, thereby allowing the 
system to remain in a dynamic, frustrated spin state to the lowest observed 
temperatures.
\end{abstract}


\date{To appear in Phys. Rev. B, scheduled issue January 1, 2004}

\maketitle 

\section{Introduction}

The term \textit{geometrical frustration}\cite{Gaulin94,Ramirez01,CanFrustProc01} indicates 
when the spins on a well-ordered lattice
interact magnetically, yet individual interactions cannot reach
their minimum energy configuration because of competing magnetic
interactions from other sites.  A simple example occurs on a two-dimensional, 
corner-shared triangular lattice with antiferromagnetic nearest-neighbor coupling, 
otherwise known as a \textit{kagom\'{e}} lattice.  The
three dimensional analogue to such a lattice is a corner-shared
tetrahedral lattice, and is realized in the pyrochlore
systems, $A_2B_2$O$_7$.

Theoretically, it has been shown
that Heisenberg, antiferromagnetic nearest-neighbor interacting
spin systems~\cite{Reimers91} do not possess a transition into an
ordered state.  
This makes corner-sharing tetrahedral spin
systems with antiferromagnetic exchange ideal candidates for
three-dimensional, low temperature spin-liquids.  A spin-liquid state
may not form if further interactions such as crystal
field anisotropy,\cite{Hertog00}, long-range exchange or dipolar
interactions perturb the classical spin models.\cite{Enjalran03}
However, since \textit{any} amount of lattice disorder can
precipitate a spin-glass phase,\cite{Edwards75} geometrically-frustrated
systems that do not magnetically order to the lowest temperatures are
extremely rare.\cite{Canals98}
This situation has inhibited
experimental studies that attempt to isolate the effects of frustration.
For instance, lattice imperfections in several geometrically frustrated systems are thought to
relieve the frustration allowing the system to freeze. N{\'e}el
order has been observed in the Jarosites,
KCr$_3$(OD)$_6$(SO$_4$)$_2$, which contains Cr$^{3+}$,
S=$\frac{3}{2}$ ions on a kagom{\'e} lattice,\cite{Lee97} due to
vacancies on the magnetic site. Site disorder in
SrCr$_{9p}$Ga$_{12-9p}$O$_{19}$,\cite{Schiffer96b} for example,
facilitates the formation of a spin glass state, as does intersite
mixing in  Gd$_3$Ga$_5$O$_{12}$.\cite{Raju92,Petrenko98}

Significantly, the pyrochlore lattice\cite{Subramanian83} appears to be 
capable of producing a spin-liquid state.  The magnetic-oxide pyrochlores have the 
chemical formula $A_2B_2$O(1)$_6$O(2) (space group $Fd \bar{3}m$).  Here, we 
focus on the case where the $A$ site is occupied by a tri-valent rare-earth ion 
with eightfold oxygen coordination, $A$O(2)$_2$O(1)$_6$,  and
the $B$ site is occupied by a tetra-valent transition metal ion with sixfold
oxygen co-ordination, $B$O(1)$_6$. The $A$ and $B$ sites
individually form infinite interpenetrating sublattices of
corner-sharing tetrahedra.  A high degree of magnetic frustration
is exhibited on this lattice when either site is occupied by 
antiferromagnetically coupled ions.  

Materials with the $A_2B_2$O$_7$ formula unit can have various structures, 
some of which can
be thought of as defected versions of the fluorite (CaF$_2$) unit cell.
The cubic fluorite cell has Ca atoms at the center of eight
F atoms at the corners of a cube.  This cell is
analogous to one eighth that of the oxide pyrochlore with a random
distribution of the metal cations, one oxygen vacancy (the 8$a$ site), and the 
positional parameter $x$ for the oxygens on the 48$f$ site equal to $3/8$.  
The rare-earth titanate pyrochlores have $x\approx0.328$.\cite{Lian03} 
Obviously, these 
structures are related, but the mixing of the metal ions in the fluorite 
structure creates multiple
exchange constants, usually resulting in a static magnetic
structure. Other materials with the stoichiometry $A_2B_2$O$_7$
include the weberites, which are closely related to the pyrochlore
structure, but only half the $B$-site ions have the sixfold oxygen
coordination, while the others are linked through only four vertices.

Despite the inherent geometrical frustration in the ideal pyrochlore
lattice, most pyrochlores order magnetically.  Tb$_2$Ti$_2$O$_7$, on the
other hand, appears to be the rare case of a pyrochlore that does not
order magnetically, exhibiting a high degree of frustration 
(the large Tb$^{3+}$ moment has a Curie-Weiss temperature of
$\sim$-19~K) down to 70~mK.\cite{Gardner99,Gingras00,Gardner01}  However, the 
pyrochlore system is also capable of producing a lattice with fairly subtle 
disorder that can precipitate a spin-glass phase, such as in the Mo-Mo 
near-neighbor pairs in Y$_2$Mo$_2$O$_7$.\cite{Booth00,Keren01}  A similar
mechanism may also precipitate the spin-glass phase in 
Tb$_2$Mo$_2$O$_7$.\cite{Gaulin92}  In fact, the 
lack of such frozen magnetic states in Tb$_2$Ti$_2$O$_7$ is somewhat 
surprising, given the results of previous structural measurements.
The crystal structure of Tb$_2$Ti$_2$O$_7$ was first investigated
by Brixner\cite{Brixner64} in 1964.  The most detailed structural study
is reported by van de Velde \textit{et al.}\cite{Velde90} 
The International Centre for Diffraction Data
quotes van de Velde's data as the standard for Tb$_2$Ti$_2$O$_7$.
Their x-ray diffraction study finds that the measured Tb$_2$Ti$_2$O$_7$ sample
has approximately 10\% disorder on the metal cation and oxygen
sublattices, and the authors conclude that Tb$_2$Ti$_2$O$_7$ is an
imperfect pyrochlore. 
The study we report below was motivated by this result and
attempts to definitively answer the question of lattice
imperfections in Tb$_2$Ti$_2$O$_7$, a significant parameter in any
model that would successfully describe the magnetic nature of
Tb$_2$Ti$_2$O$_7$.

The present work reports 
neutron powder diffraction (NPD) measurements of the 
average
structure, and carefully considers the thermal dependence of the mean-squared
displacement parameters as a means of separating the non-Debye (presumably
static) component of any positional disorder.  In addition, the 
x-ray absorption fine-structure (XAFS) technique is used to check for
disorder in the local environment around Tb and Ti atoms, and in particular
for the kind of $B$-site correlated disorder that occurs in Y$_2$Mo$_2$O$_7$.
Finally, the results from the two techniques are combined to determine 
to what degree correlations exist between several atom pairs.

The rest of this paper is organized as follows:  Sample
preparation and experimental details are in Sec. \ref{exp_det}.
Data analysis procedures and results are reported in Sec. \ref{results}.
Consequences of these results, comparisons to previous measurements, and
further discussion are in Sec. \ref{discussion}, and the final conclusions are 
in Sec. \ref{conclusions}.

\begin{figure}[t]
\includegraphics[height=3.2in,angle=90]{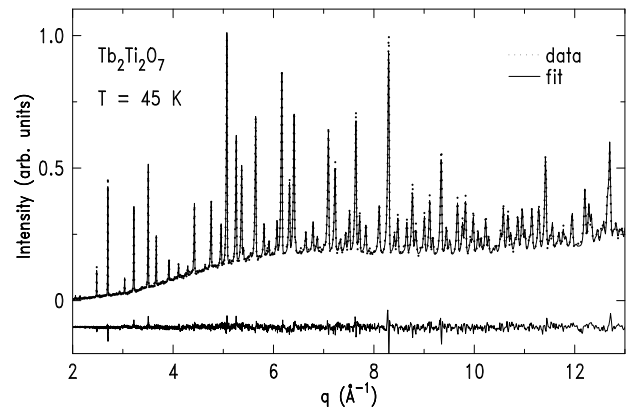}
\caption{Neutron diffraction data collected at IPNS from powder
Tb$_2$Ti$_2$O$_7$ at 45 K as a function of the momentum transfer $q$.  The solid
curve is the best fit from the Rietveld refinement with
the $Fd \bar{3}m$ structure. The 
bottom curve shows the difference between the measured and calculated
intensities.} 
\label{npd}
\end{figure}

\section{Experimental Details}
\label{exp_det}

Polycrystalline Tb$_2$Ti$_2$O$_7$ was prepared by a conventional
solid state reaction in the same manner as in Ref. \onlinecite{Gardner99},
namely, by firing stoichiometric amounts of Tb$_4$O$_7$ and TiO$_2$ at 
1350$^\circ$ for several days with intermittent grindings to ensure a complete 
reaction.
The neutron diffraction measurements were performed on the time-of-flight 
Special Environment Powder Diffractometer at the Intense Pulsed Neutron Source (IPNS) 
at Argonne National Laboratory and on the C2 diffractometer at the NRU
reactor at Chalk River Laboratories. 
The ground powder was packed in a vanadium can and
cooled in a displex (IPNS) or a helium-bath cryostat (NRU). 
A Si(113) crystal was used at the NRU reactor
to produce a monochromatic neutron beam with a wavelength of 1.3283 \AA.
The large difference in the scattering lengths, in
particular the negative scattering length of titanium, creates a large
contrast with the terbium scattering.  This contrast allows for
an accurate determination of the crystal structure properties, and especially 
for any possible cation antisite disorder.  

For XAFS measurements,
the polycrystalline Tb$_2$Ti$_2$O$_7$ sample was re-ground and passed through
a 20~$\mu$m sieve. The sieved powder was uniformly
distributed over adhesive tape, cut into strips and stacked to
obtain absorption edge steps of 0.8 absorption lengths at the Ti $K$ edge and
1.0 absorption lengths at the Tb $L_\textrm{III}$ edge. Samples were then
placed into a LHe-flow cryostat.  Transmission XAFS data were collected
at the Stanford Synchrotron Radiation
Laboratory (SSRL).  Tb $L_\textrm{III}$-edge
(7515 eV) data were collected on beamline 4-3 
with a 1/2-tuned Si(220) double-crystal monochromator.  Ti $K$-edge (4965 eV)
data were collected on beamline 2-3 with a 1/2-tuned Si(111) double-crystal
monochromator.

\begin{figure}[t]
\includegraphics[width=3.2in,angle=0]{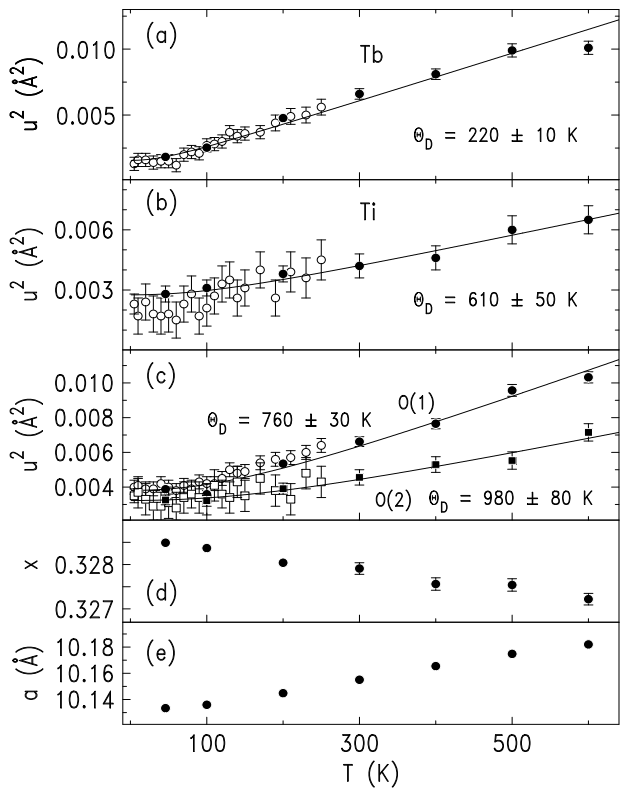}
\caption{Mean-squared displacement parameters, $u^2$'s,  for the 4
crystallographic sites of Tb$_2$Ti$_2$O$_7$ as a function
of temperature are shown in panels (a)-(c). The data were collected at the 
IPNS (closed symbol) and 
the NRU reactor (open symbol).  Solid lines are best fits using a Debye model 
(see Table \ref{tableNPD}). Panels (d) and (e) show the temperature dependence 
of the positional parameter $x$ of the O(1) oxygen (48$f$ site) and the
lattice parameter $a$, respectively.} 
\label{u2x}
\end{figure}

\section{Results}
\label{results}

The rare earth oxide pyrochlores crystallize in the space group
$Fd \bar{3}m$ as mentioned above, with 4 crystallographically
independent sites ($A^{3+}$ in position 16$d$ at ($\frac{1}{2},
\frac{1}{2}, \frac{1}{2}$), $B^{4+}$ in 16$c$ at (0, 0, 0), O(1) in 48$f$
at ($x, \frac{1}{8}, \frac{1}{8}$) and O(2) in 8$b$ at
($\frac{3}{8}, \frac{3}{8}, \frac{3}{8}$)\cite{Subramanian83}).  The position
of the 48$f$ site is defined by only one variable, $x$, which is
sensitive to lattice effects, including the absolute size and
the relative difference in size of the metal ions, as well as to various forms
of disorder.\cite{Minervini02}  

\begin{table}
\caption{NPD fit results to IPNS data from Tb$_2$Ti$_2$O$_7$ using the
$Fd\bar{3}m$ space group.  Reported values are from fits at 45~K,
except for the Debye fit results, which are over the full temperature range of the
data (45 - 600~K).
Static disorders were obtained with the Debye model fit shown in
Fig.~\ref{u2x}.  All site occupancies are fixed at unity. }
\begin{ruledtabular}
\begin{tabular}{llllll}
General fit characteristics: &&&&& \\
Banks included          &&      \multicolumn{2}{l}{$\pm$ 145$^\circ$, $\pm 90^\circ$} && \\
Total data points       &&      8792 &&&\\
Total measured reflections &&   362&&&  \\
\# of variables          &&    \multicolumn{3}{l}{9+4 for background} &\\
\\
reduced $\chi^2$        &1.802 &&&&\\
$Rp$(\%)                &6.90 &&&&\\
$wRp$(\%)               &4.51 &&&&\\
\end{tabular}
\begin{tabular}{lllllll}
Atom & $x$ & $y$ & $z$ & $u_\textrm{iso}^2 (\AA)$ & $\Theta_\textrm{D}$ (K)
& $u_\textrm{static}^2$(\AA$^2$) \\
\colrule
Tb & 1/2 & 1/2 & 1/2 & 0.0018(2) & 220(10)  & 0.0006(7) \\
Ti & 0 & 0 & 0 & 0.0028(4) & 610(50) & 0.0015(12) \\
O(1) & 0.3285(1) & 1/8 & 1/8 & 0.0039(2) & 760(30) & 0.0010(12) \\
O(2) & 3/8 & 3/8 & 3/8 & 0.0033(3) & 980(80) & 0.0008(12) \\
\\
$a_0$ & \multicolumn{2}{l}{10.13315(12) \AA} &&&&\\
\\
\multicolumn{2}{l}{Tb$-$O(2)$-$Tb} & \multicolumn{2}{l}{109.47(1)$^\circ$} &&&\\
\multicolumn{2}{l}{Tb$-$O(1)$-$Tb} & \multicolumn{2}{l}{106.41(3)$^\circ$} &&&\\
\multicolumn{2}{l}{Tb$-$O(1)$-$Ti} & \multicolumn{2}{l}{106.41(2)$^\circ$} &&&\\
\multicolumn{2}{l}{Ti$-$O(1)$-$Ti} & \multicolumn{2}{l}{132.12(5)$^\circ$} &&&\\
\end{tabular}
\end{ruledtabular}
\label{tableNPD}
\end{table}

The neutron diffraction data were analyzed with a Rietveld
method using the General Structure Analysis System (GSAS) refinement
package.\cite{gsas} 
Figure\ \ref{npd} shows the 45~K data 
together with the best refinement and the residual, shown at the bottom.
No additional peaks, associated with either a second phase or a
magnetic unit cell, were observed over the whole temperature range from
4.5-600~K. 
However as previously
reported,\cite{Gardner99} an anomalous background due to short range
magnetic correlations was observed at the lowest temperatures. 
The atomic displacement parameters of all atoms were assumed to be
isotropic during the fitting process since no improvement in the
refinement was obtained by using anisotropic parameters.  This is in
contrast to Y$_2$Mo$_2$O$_7$, where anisotropic parameters for the $A$ site
were deemed necessary.\cite{Reimers88}
The results of the refinement are summarized in Table~\ref{tableNPD}.
The possibility of antisite disorder (site interchange between Tb and Ti
atoms) was also considered, such as could occur if the system had a propensity
toward a fluorite-like structure.\cite{Wang99,Chartier02}  We find no evidence 
for any such disorder, and place an upper limit of 2\% on metal-site mixing. 
Similarly, we performed fits allowing the oxygen occupancy fractions to vary, 
including the possibility of oxygen on the nominally vacant 8$a$ site.  These
measurements provide no evidence of excess or deficient oxygen compared to the
nominal pyrochlore structure within an upper limit of 2\%.
Both of these limits are considerably lower than the disorder reported by 
van de Velde \textit{et al.}\cite{Velde90}

Figure\ \ref{u2x} shows the mean-squared displacements, $u^2$, for all
four crystallographic sites and the positional parameter, $x$, as a
function of temperature.  No structural anomalies, within
uncertainty, were observed. The $u^2$'s for each site were fit with a thermal 
model, including a temperature-independent offset:  
\begin{equation}
        u_\textrm{meas}^{2}(T) = u_\textrm{static}^{2} + u_\textrm{thermal}^2(T).
\label{u_model}
\end{equation}
The temperature-dependent part of the mean-squared displacement parameters,
$u_\textrm{thermal}^2(T)$, is given in general by a collection of quantum 
harmonic oscillators:
\begin{equation}
u_\textrm{thermal}^2(T) =\frac{\hbar}{2 m} \int_0^\infty \rho(\omega)
coth\left(\frac{\hbar \omega}{2 k_\textrm{B} T} \right)
\frac{d\omega}{\omega}
\label{debye_model}
\end{equation}
where $m$ is the atomic mass and $\rho$ is the phonon density of states 
at the given site.
In the Debye model, the density of states is empty above a cutoff frequency,
$\omega_\textrm{D}$, and quadratic below it:  
$\rho = 3\omega^{2}/\omega_\textrm{D}^{3}$.  The Debye temperature 
is then just $\theta_\textrm{D}=\hbar\omega_\textrm{D}/k_\textrm{B}$.
Note that zero-point motion exists such that even in the absence of a static
offset, $\mu_\textrm{thermal}^2$ is not zero at $T$=0, and is, in fact:
\begin{equation*}
u_\textrm{thermal}^2(T=0) =\frac{3\hbar^2}{4 m k_\textrm{B} \Theta_\textrm{D}}.
\end{equation*}
The temperature-independent offset in addition to this zero-point motion, 
$u_\textrm{static}^2$, when
observed, can be a strong indication of the presence of static (non-thermal)
disorder in a system.  The temperature dependence of the $u^2$
parameters is described well by the Debye model as
shown by the fits in Fig.~\ref{u2x}.  In Table~\ref{tableNPD}
we summarize these fits.  The $u^2_\textrm{static}$ parameters are negligibly 
small at all atomic sites, and the Debye 
temperatures are appropriate for the elements they represent.

\begin{figure}
\includegraphics[width=3.2in,angle=0]{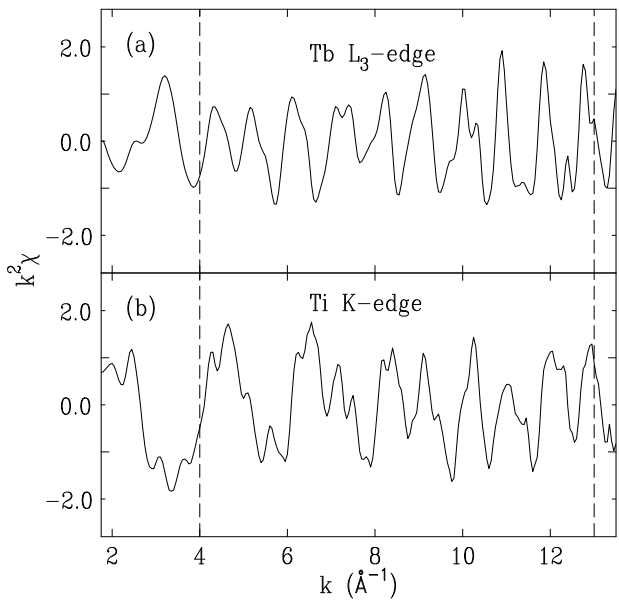}
\caption{XAFS data at 20 K measured above the (a) Tb $L_3$-edge and (b) Ti $K$-edge 
as a function of photoelectron wave number, $k$.}
\label{kspace}
\end{figure}

Investigations of the local environment around the metal ions were
performed using temperature dependent XAFS above the Tb
$L_\textrm{III}$ edge and the Ti $K$ edge.
XAFS oscillations occur above a core-level x-ray absorption edge, and 
are due to the interaction between the outgoing photoelectron from the
absorbing atom and the component of this photoelectron that is back-scattered 
by neighboring atoms.  The oscillations are periodic in the pair distance,
$r$, and the photoelectron
wave number, $k = \sqrt{2m(E - E_{\rm 0})} / \hbar$, where $E$ is the
incident x-ray energy and $E_{\rm 0}$ is the energy at the absorption edge.
XAFS analysis gives information about the radial distance distribution around
the absorbing atomic species.  In particular, pair distances, $r_i$, and 
pair-distance distribution widths, $\sigma_i^2$, can be obtained for the first 
few atomic shells, $i$, out to about 5 \AA.
These XAFS data were analyzed using the UWXAFS package \cite{UWXAFS}. 
First, the XAFS function $\chi=\mu/\mu_0-1$ is extracted after determining 
the atomic background absorption, $\mu_0$, obtained with the UWXAFS program 
AUTOBK by passing a cubic spline through the absorption data, $\mu$.
The $k$-space XAFS data for both measured edges are
shown in Fig.\ \ref{kspace}.  Multiple scans were
obtained at each edge and temperature, with excellent reproducibility.

\begin{figure}[t]
\includegraphics[width=3.2in,angle=0]{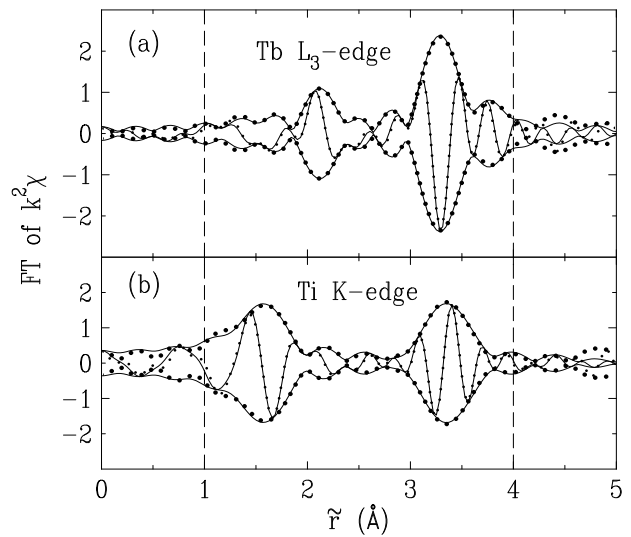}
\caption{Magnitude of the Fourier transformed (FT) XAFS data from 
Fig. \ref{kspace} (thick-dotted lines) and the real part of the XAFS 
(thin-dotted lines)
as function of distance from the probe atoms, (a) Tb $L_3$-edge and
(b) Ti $K$-edge. XAFS data in the range of 4.0 $-$ 13.0 
\AA$^{-1}$ are used
for the Fourier transform with a 0.5 \AA$^{-1}$ wide Hanning window.
 Solid lines are the best fits and 
the vertical lines indicate the fit regions.}
\label{rspace}
\end{figure}

\begin{figure}[t]
\includegraphics[width=3.2in,angle=0]{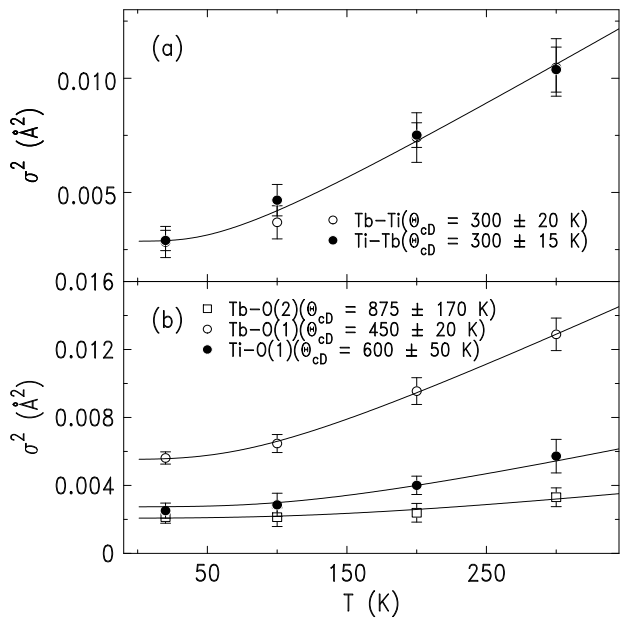}
\caption{Debye-Waller factor($\sigma^2$) as a function of temperature. 
(a) Tb$-$Ti and Ti$-$Tb pairs, and (b) Tb$-$O(1), Tb$-$O(2) and Ti$-$O(1) pairs 
determined at Tb $L_3$-edge and Ti $K$-edge.  Solid lines are the best fits 
with the correlated-Debye model (see Table \ref{tableXAFS}). Ti-Ti pairs are
omitted for clarity.}
\label{sigs}
\end{figure}

Figure\ \ref{rspace} shows the magnitude and the real part of the Fourier 
transformed XAFS for both edges.
Note that the peaks are shifted on the $\tilde{r}$ axis from their true bond 
lengths due to
the phase shift of the back-scattered photoelectron.  This phase shift, as well
as other details of the complex back-scattering function, differ for each
atomic species, although they are well approximated by the theoretical
treatment provided by the FEFF8 code.\cite{FEFF8}  Detailed fits are
therefore necessary to obtain quantitative information.
Solid lines show the best fit of a multiple shell model that assumes full
occupancy of each of the sites, but allows the $r_i$ and the $\sigma_i^2$ 
parameters for each shell to vary, with some exceptions.  In
particular, the Tb-Ti and Tb-Tb pairs near 3.60 \AA\ are constrained to
have the same pair distance, as are the Ti-Ti and Ti-Tb pair distances.
The fit results are given in Table \ref{tableXAFS} and 
Fig. \ref{sigs}. 
Additional details of the XAFS analysis and the determination 
of the uncertainty of parameters have been given elsewhere.\cite{UWXAFS,Han02}

\begin{table*}
\caption{XAFS fit results at 20 K.  The overall amplitude reduction factors, 
$S_0^2$, were determined to be 0.96(9) for the Tb $L_\textrm{III}$-edge fits 
and 0.80(6) for the Ti $K$-edge fits.  The number of fit parameters (10 and 
13, respectively), is much less than the number of independent data points 
determined by Stern's rule\cite{Stern93} (19 for each). $N$ is the coordination
number for the fully occupied model.  Bond lengths of atomic pairs 
determined using XAFS ($r$) are compared to the NPD results 
($r_\textrm{NPD}$) at 45~K. 
The correlated-Debye temperatures, $\Theta_\textrm{cD}$, and the 
temperature-independent offsets, $\sigma_{static}^2$, were determined by 
fitting with a correlated-Debye model\cite{Beni76,Crozier88} over the measured 
temperature range (20-300~K).  The correlation parameter $\phi$ is given by
Eq. \ref{corr}. }
\begin{ruledtabular}
\begin{tabular}{ccccccc}
Atomic pair & $N$ & $r$(\AA) & $r_\textrm{NPD}$(\AA) & $\Theta_\textrm{cD}$(K)
& $\sigma^2_{static}$(\AA$^2$) & $\phi$ \\
\colrule
Tb$-$O(2) & 2 & 2.199(6) & 2.1939(6) & 875(170) & -0.0002(7) &  0.61(7) \\ 
Tb$-$O(1) & 6 & 2.490(9) & 2.4958(7) & 450(20) & 0.0004(5) & 0.18(7)\\ 
Tb$-$Ti$^1$ & 6 & 3.598(6) & 3.583(1) & 300(20) & -0.0004(8) & 0.40(15)\\ 
Tb$-$Tb$^1$ & 6 & 3.598(6) & 3.583(1) & 269(11) & -0.0003(3) &  0.65(9)\\
\\
Ti$-$O(1) & 6 & 1.985(7) & 1.9699(5) & 600(50) & -0.0009(10) & 0.60(7) \\ 
Ti$-$Ti$^2$ & 6  & 3.596(7) & 3.583(1) & 500(60) & -0.0005(7) & 0.55(8)\\ 
Ti$-$Tb$^2$ & 6 & 3.596(7) & 3.583(1) & 300(15) & -0.0003(5) & 0.38(10)\\
\end{tabular}
\end{ruledtabular}
$^1$ Tb$-$Ti \& Tb$-$Tb bond length constrained together \\
$^2$ Ti$-$Ti \& Ti$-$Tb bond length constrained together \\
\label{tableXAFS}
\end{table*}

The $\sigma^2$ data were analyzed analogously to the NPD $u^2$ parameters
described above, except that a 
\textit{correlated}-Debye model\cite{Beni76,Crozier88} was used to 
account for correlated motion/displacements of the atom pairs.  This model for 
$\sigma^2$ is the 
same as the Debye model (Eq. \ref{u_model} and \ref{debye_model}) except that 
a correlated-Debye temperature, $\Theta_\textrm{cD}$,
is used to differentiate from the regular Debye model, the reduced mass, $\mu$, of the atom pairs is
used, and the phonon density of states for a given pair at distance $R$ is given by
\begin{equation}
\rho = \frac{3 \omega^{2}}{\omega_\textrm{D}^{3}} \left[ 1 -
\frac{sin(\omega R/c)}{\omega R/c}\right],
\label{dos}
\end{equation}
where $\omega_\textrm{D}$ is the usual Debye cut-off frequency and 
$c=\omega_\textrm{D}/k_\textrm{D}$. The Debye wave vector is given by
$k_\textrm{D}=(6 \pi^2/V)^{1/3}$, where $V$ is the mean
volume per atom in the material.
The expression in brackets of Eq. \ref{dos}
takes into account the correlated motion of the atom pair.
This model fits the temperature dependence of the measured $\sigma^2$'s
well (Fig. \ref{sigs}) with only 
negligible static displacements, $\sigma_\textrm{static}^2$'s, as shown in 
Table \ref{tableXAFS}.  
This lack of pair-distance disorder agrees well 
with the lack of positional disorder determined from the NPD results above. 

A Tb/Ti site interchange model was also considered for these XAFS data.  
However, XAFS is insensitive to such site interchange 
in this system because correlations in the fits between the (nominally same) 
Tb-Tb and Tb-Ti pair distances (as well as the Ti-Tb, Ti-Ti distances), 
their $\sigma^2$'s, and the alleged interchange render the results 
inconclusive.

\section{Discussion}
\label{discussion}

The primary result of the above measurements is that the Tb$_2$Ti$_2$O$_7$
lattice is extremely well ordered, both on average (NPD) and locally (XAFS).
This result is supported by the high quality of the fits, the low estimated
errors, and the temperature dependence of the mean-squared displacement
parameters.  These data thus strongly support the conclusion made by previous
authors that Tb$_2$Ti$_2$O$_7$ is an excellent laboratory to study the effects
of geometric frustration in the absence of lattice disorder.

The discrepancy between our results and those of 
van de Velde \textit{et al.}\cite{Velde90} 
must be attributable to either the different techniques used to measure
the structure or to produce the sample.  Since powder x-ray diffraction
should be able to distinguish the cations, we must 
assume that the synthesis method used in van de Velde \textit{et al.},
namely, preparing  
Tb$_2$Ti$_2$O$_7$ by the citrate method from
TiO(OH)$_2$ and a solution of Tb$_4$O$_7$ in nitric acid, results
in an imperfect pyrochlore lattice.  This situation is not inconceivable,
since considerably lower
temperatures and shorter times are used in this method compared to the
solid state reaction method used here.\cite{Gardner99}

Very recently, Lian \textit{et al.}\cite{Lian03} reported an x-ray
diffraction study of the entire
RE$_2$Ti$_2$O$_7$ (RE=rare-earth) series.  Although they do not focus on 
Tb$_2$Ti$_2$O$_7$ or on the detailed search for lattice disorder presented here,
they report a room-temperature $x$ parameter for the oxygen 48$f$ site of 
0.3281(5), which is consistent with our measurement of 0.3279(1), but not
with the value of 0.317(2)\cite{pyro_note} found by 
van de Velde \textit{et al.}\cite{Velde90} A value of $x\approx0.328$
is, however, consistent with that found in the RE$_2$Ti$_2$O$_7$ 
series,\cite{Lian03} while 
a value less than 0.32 is a gross deviation.  As noted earlier, this
parameter is very sensitive to lattice distortions, which could be due to
antisite disorder, oxygen excess or deficiency, etc.  In fact, 
Minervini \textit{et al.}\cite{Minervini02} find that the theoretical ideal
value for $x$ given no lattice disorder is $\sim$0.3278, in excellent
agreement with our measurements and those of Lian \textit{et al.}  This agreement,
together with the fact that
Lian \textit{et al.} used a flux-growth synthesis technique and obtained
similar results to those presented here, is therefore consistent with our 
interpretation that the main reason van de Velde \textit{et al.}
found Tb$_2$Ti$_2$O$_7$ to be an imperfect pyrochlore is the synthesis method,
not the measurement technique, nor is the disorder generic to the material 
itself.

There are a few interesting points to note about the measured structure and its
vibrational properties.  One is that the single-site Debye temperatures
(from NPD) do not scale as expected in a Debye solid, that is, by the
square root of the ratio of their masses.  In this way, one sees that the Tb
atomic vibrations are governed by much softer phonons that the Ti atoms 
[$\sqrt{m_\textrm{Ti}/m_\textrm{Tb}} \Theta_\textrm{D}(\textrm{Ti}) \approx330$ K, while 
$\Theta_\textrm{D}$(Tb)=220~K].  

Another aspect to consider is the correlated displacements of the atom pairs.
Since we have both neutron diffraction and XAFS measurements, 
we can extract the correlations by combining the results from both techniques.
The instantaneous distance ($\Delta \vec{r}_{AB}$) 
between two atoms ($A, B$) is equal to the difference of the instantaneous 
displacements of the individual sites, $\Delta \vec{r}_A - \Delta \vec{r}_B$.
Taking $\sigma_{AB}^2=<\Delta \vec{r}_{AB}^{~2}>$, $u_X^2=<\Delta \vec{r}_{X}^{~2}>$
and $u_A u_B \phi = <\Delta \vec{r}_A \cdot \Delta \vec{r}_B>$, we then have
\begin{equation}
\sigma_{AB}^2 = u_A^2 + u_B^2 - 2\phi u_A u_B.
\label{corr}
\end{equation}
The correlation coefficient, $\phi$, describes the motion or 
displacements of an atom relative to the other atom, taking the values +1 for 
two atoms moving together in the same direction, -1 for the atoms moving in 
fully opposite directions, and 0 for uncorrelated motion/displacements. 
The $\phi$'s of Tb$_2$Ti$_2$O$_7$
at the lowest temperature extracted from the
XAFS and diffraction measurements are summarized in Table \ref{tableXAFS}.
The motions of the two O(2) oxygens located at $\sim$2.2 \AA\ from the Tb atom and
the six O(1) oxygens located at $\sim$2.0 \AA\ from Ti atom  
are correlated in the motions to Tb and Ti, respectively, 
with $\phi$ of $\sim$60\%.
It is somewhat surprising that their motions are not more highly correlated
with such short bond lengths, especially compared to, say, the copper-oxide
superconductors.\cite{Booth96} More interestingly, the Tb-O(1) pairs
are nearly uncorrelated in their motions, with a $\phi$ of only 0.2.  Therefore,
the motions of the O(1) atoms are governed mainly by the Ti sublattice, and
not by the Tb sublattice.  This result is consistent with the Tb sites 
exhibiting softer phonons than expected from the Ti vibrations.

Given these vibrational properties, one should consider the charge distribution 
in the various ligands.  One simple method is to determine whether the 
bond-valence sums\cite{Brown85} for the Tb and Ti oxygen environments correspond
to the expected values of +3.0 and +4.0, respectively.  Indeed, we obtain +3.01
and +3.95 from the NPD data, indicating that the charge is distributed as 
expected, given the
measured coordination and bond lengths.  Therefore, the measured lack of 
correlation between Tb and O(1) atoms and the soft Tb phonons are most likely a 
generic feature of a well-ordered pyrochlore lattice, such as 
Tl$_2$Mn$_2$O$_7$, or any of the other ferromagnetic pyrochlores.

A natural question to ask then is ``Why is Tb$_2$Ti$_2$O$_7$ structurally so
well ordered?''  To this end, we can only speculate as to why, for instance, 
Y$_2$Mo$_2$O$_7$ exhibits distortions while Tb$_2$Ti$_2$O$_7$ does not.  
Ionic size arguments are marginally fruitful, since 
the ionic-size ratio\cite{Shannon76} for Tb$_2$Ti$_2$O$_7$ ($r_A/r_B=1.72$) is 
rather different than for Y$_2$Mo$_2$O$_7$ ($r_A/r_B=1.57$).  However, both of 
these values fall well within the range 1.46 to 1.80 typical of the 
pyrochlores.\cite{Subramanian83}
On the other hand, the negligibly small amount of static disorder on all atomic 
sites and between all near-neighbor pairs might suggest that
the interaction of corner sharing spins is not strong enough
to distort the atomic sites, as has been conjectured for 
Y$_2$Mo$_2$O$_7$.\cite{Booth00,Keren01}  This explanation is supported by the
relatively small Curie-Weiss temperature of $\sim-19$~K in Tb$_2$Ti$_2$O$_7$, 
compared to $\sim-200$~K in Y$_2$Mo$_2$O$_7$.\cite{Gingras97}

\section{Conclusion}
\label{conclusions}

We have examined the bulk and local structure of the spin liquid
pyrochlore Tb$_2$Ti$_2$O$_7$ by powder neutron diffraction and
XAFS at the Tb $L_\textrm{III}$ and Ti $K$ edges.  These data indicate a 
high degree of crystalline order, suggesting that this compound forms a perfect
pyrochlore lattice with no intersite mixing or anion disorder.
Furthermore, analysis of the temperature dependence of the atomic
displacement parameters and mean-squared displacements shows that
the static lattice disorder is negligible.  Analysis of the Debye
temperatures and the correlations between various site displacements
show that, vibrationally, the Tb and Ti sublattices are partially decoupled, 
and that the more plentiful O(1) atoms are more tightly bound to the Ti sites.  
This compound, with its near absence of local and bulk structural disorder and 
antiferromagnetic 
nearest-neighbor interactions,\cite{Gardner99,Gingras97} is therefore an 
appropriate system to study the elusive 3-dimensional, low-temperature, spin 
liquid. 

\acknowledgments

It is a pleasure to acknowledge A. Cull and I. Swainson for their
assistance with the experiments at Chalk River, and S. Short for
her assistance with the experiments at the Intense Pulsed Neutron Source (IPNS).
Work at Lawrence Berkeley National Laboratory was supported by the Director,
Office of 
Science, U.S. Department of Energy (DOE) under Contract No. DE-AC03-76SF00098.
Work at Brookhaven National Laboratory was supported by the Division of Material 
Sciences, DOE, under contract DE-AC02-98CH10886.  Neutron diffraction data were 
collected at the Chalk River Laboratories NRU reactor and at the IPNS, which is 
funded by the DOE, Office of Basic Energy Sciences (OBES), Materials Sciences, 
under Contract W-31-109-Eng-38.  X-ray absorption data were collected at the 
Stanford Synchrotron Radiation Laboratory, a national user facility operated by 
Stanford University for the DOE, OBES.

\bibliographystyle{apsrev}
\bibliography{/home/hahn/chbooth/papers/bib/bibli}

\begin{thebibliography}{39}
\bibitem[{pyr()}]{pyro_note}
\bibinfo{note}{Some diffraction studies of pyrochlore structures place the $A$
  atoms on the 16$c$ site and the $B$ atoms on the 16$d$ site, in contrast to
  the present work. In this case, the $x$-parameter for the 48$f$ sites is
  related by taking 0.75-$x$. Also, the O(2) atoms are then on the 8$a$ sites
  while the 8$b$ sites are vacant, rather than the other way around.}
\end{thebibliography}

\end{document}